%% file: sample-1col.tex
\documentclass[]{ceurart}
\usepackage{listings}
\usepackage{kotex}
\usepackage[most]{tcolorbox}
\usepackage{float}
\usepackage{algorithm}
\usepackage{algorithmic}
\usepackage[boxed,lined,algosection,algo2e]{algorithm2e}
\usepackage{amsmath,amssymb}
\usepackage{amsmath, amssymb}    % \vee, \bigcup 등 사용
\usepackage{xcolor}
\tcbuselibrary{listings, breakable}
\lstdefinestyle{mypython}{
    language=Python,
    basicstyle=\small\ttfamily,
    breaklines=true,
    breakatwhitespace=false,
    showstringspaces=false,
    keepspaces=true,
    columns=fullflexible,
    upquote=true,
    keywordstyle=\color{blue!70!black}\bfseries,
    stringstyle=\color{teal!60!black},
    commentstyle=\color{gray!80},
    tabsize=4
}

\begin{document}
\copyrightyear{2025}
\copyrightclause{Copyright for this paper by its authors. Use permitted under Creative Commons License Attribution 4.0 International (CC BY 4.0).}

\conference{Proceedings of RDGENAI '25: First International Workshop on Retrieval-Driven Generative AI at CIKM 2025, November 14, 2025, Coex, Seoul, Republic of Korea}

\title{SHRAG: A Framework for Combining Human-Inspired Search with RAG}

\author[1]{Hyunseok Ryu}[%
orcid=0009-0000-4260-6287,
email=omnyx2@gmail.com,
url=https://www.linkedin.com/in/hyunseok-ryu-376534170/,
]
\cormark[1]
\fnmark[1]
\address[1]{Department of AI Convergence, Gwangju Institute of Science and Technology(GIST) 123 Cheomdangwagi-ro(Oryong-dong),Buk-gu, Gwangju 61005, Republic of Korea}

\author[1]{Wonjune Shin}[%
email=shinwj10@gm.gist.ac.kr,
]
\fnmark[1]
\author[2]{Hyun Park}[%
email= andrew630300@gmail.com,
]
\address[2]{College of Information Computing
Gwangju Institute of Science and Technology(GIST) 123 Cheomdangwagi-ro(Oryong-dong),Buk-gu, Gwangju 61005, Republic of Korea}

\cortext[1]{Corresponding author.}
\fntext[1]{These authors contributed equally.}

\begin{abstract}
Retrieval-Augmented Generation (RAG) is gaining recognition as one of the key technological axes for next-generation information retrieval, owing to its ability to mitigate the hallucination phenomenon in Large Language Models (LLMs) and effectively incorporate up-to-date information. However, specialized expertise is necessary to construct a high-quality retrieval system independently; moreover, RAG demonstrates relatively slower processing speeds compared to conventional pure retrieval systems because it involves both retrieval and generation stages.
Accordingly, this study proposes SHRAG, a novel framework designed to facilitate the seamless integration of Information Retrieval and RAG while simultaneously securing precise retrieval performance. SHRAG utilizes a Large Language Model as a Query Strategist to automatically transform unstructured natural language queries into logically structured search queries, subsequently performing Boolean retrieval to emulate the search process of an expert human searcher. Furthermore, it incorporates multilingual query expansion and a multilingual embedding model, enabling it to perform efficient cross-lingual question answering within the multilingual dataset environment of the ScienceON Challenge.
Experimental results demonstrate that the proposed method, combining logical retrieval capabilities and generative reasoning, can significantly enhance the accuracy and reliability of RAG systems. Furthermore, SHRAG moves beyond conventional document-centric retrieval methods, presenting the potential for a new search paradigm capable of providing direct and reliable responses to queries.
\end{abstract}


\maketitle

\section{Introduction}
Retrieval-Augmented Generation (RAG) systems are recently emerging as a key paradigm capable of overcoming the limitations of traditional search systems by dynamically integrating the inherent knowledge of pre-trained Large Language Models (LLMs) with external information sources \cite{gao2024retrievalaugmentedgenerationlargelanguage}. This approach is being evaluated as an effective method for mitigating the hallucination phenomenon in LLMs, ensuring information freshness, and enhancing domain specificity \cite{lewis2021retrievalaugmentedgenerationknowledgeintensivenlp}.

However, several constraints exist in practical application environments. First, LLM-based generators incur high computational costs and inference latency; furthermore, the process of selecting and combining appropriate retrievers for a given context is complex. In particular, because Dense Retrievers rely on semantic similarity, cases have been reported where their performance is inferior to traditional Sparse Retrievers (e.g., BM25) in tasks requiring exact keyword matching or specialized terminology processing \cite{thakur2021beirheterogenousbenchmarkzeroshot}. To compensate for this, hybrid retrieval approaches combining Dense and Sparse methods have been proposed, yet specialized tuning is still required for domain optimization.

To experimentally explore these practical limitations, the Korea Institute of Science and Technology Information (KISTI) hosted the ScienceON AI Challenge. This competition aimed to integrate KISTI's academic search platform with RAG systems, with the objective of performing accurate document retrieval, coherent context combination, and reliable answer generation within limited resources (50GB GPU). The competition dataset includes 50 real scientific queries, corresponding reference papers, and expert-written annotations; its primary characteristic is the requirement to perform cross-lingual retrieval in a multilingual environment where Korean and English are mixed \cite{ref15}.

Considering the structural characteristics of the ScienceON AI Challenge, this study addresses three key problems: 
\begin{itemize}
\item  Automatically generating logical and structured search queries suitable for search engines from unstructured natural language queries.
\item Expanding queries multilingually and effectively exploring relevance with documents in different languages through cross-lingual retrieval.
\item Generating contextually coherent resposes from retrieved information to enhance the reliability of the RAG system.
\end{itemize}

To address these requirements, we propose a plug-and-play framework named SHRAG (Search like human with RAG). SHRAG possesses a structure that organically combines the logical retrieval functions of traditional search systems with the generative reasoning capabilities of LLMs. Additionally, SHRAG performs cross-lingual retrieval, which sophisticatedly matches queries and documents in different languages, by utilizing multilingual query expansion and a multilingual embedding model. Through this, it aims to overcome the limitations of single language based RAG and enable its use in real-world global academic data environments. This structure minimizes the use of LLMs to the query transformation stage, thereby reducing computational costs and latency, while simultaneously securing both cost-effectiveness and practicality by leveraging existing powerful search systems to their fullest extent. Furthermore, SHRAG moves beyond search results that merely list documents, presenting a new search framework paradigm that can reliably construct a list of sources for a query and provide natural language responses (direct answers) in a Plug and Play manner. The contributions we present through this are as follows:
\begin{itemize}
\item We propose an integrated design direction for cost-effective and practical search and RAG in multilingual academic sites.
\item We present the first case of utilizing an LLM as a Boolean Query Generator, applying it to existing search systems in a Plug and Play manner, and integrating this into RAG.
\end{itemize}

\section{Preliminaries}
We will briefly summarize the existing research in this field and the background knowledge necessary for this study. 
\subsection{Retrieval-Augmented Generation (RAG)}
Since RAG was first proposed to address hallucination and long-term memory, it has become the most widely utilized method in Information Retrieval leveraging LLMs, establishing itself as a core methodology that has elevated Information Retrieval performance \cite{lewis2021retrievalaugmentedgenerationknowledgeintensivenlp}. Early RAG possessed a simple structure where a Retriever fetched documents based on a Vector DB and a Generator produced answers based on them; recently, however, it is advancing rapidly through the application of various techniques such as multi-hop RAG, which breaks down complex questions into multiple steps, Self-corrective RAG, which assesses the relevance of retrieved documents and regenerates queries, and Modular RAG. Furthermore, internally, it is evolving by hybridizing Dense and Sparse Retrievers, the core algorithms for document fetching \cite{ref2}.

\subsection{Dense and Sparse Retrieval}
The field of information retrieval has traditionally been dominated by keyword-based Sparse Retrieval methods, such as TF-IDF and BM25. These methods are fast and robust in keyword matching, but they have the drawback of failing to capture synonyms or contextual meaning. Conversely, Dense Retrieval, which utilizes models like BERT, embeds text into a high-dimensional vector space and retrieves documents based on semantic similarity. While this approach shows strength in contextual understanding, it exhibits vulnerability to lexical mismatch and Out-of-Distribution queries, as previously mentioned \cite{thakur2021beirheterogenousbenchmarkzeroshot}. Recently, hybrid search, which combines these two, has been recording state of the art performance and is establishing itself as the standard approach.

\subsection{Multilingual Embedding}
Embedding is a core algorithm that vectorizes natural language sentences, enabling computers to compare sentence similarity. Early Embedding in RAG researches have shown good performance for English, but have demonstrated poor performance for multilingual contexts \cite{ref4}. However, various attempts are recently underway to achieve good performance  even with sentences that mix multiple languages. Models like mGTE and SNOWFLAKE are producing good results on multilingual benchmarks such as MMTEB \cite{ref5, ref6, ref7, ref8}.

\subsection{Single-Hop and Multi-Hop}
Multi-hop reasoning, as opposed to single-hop retrieval, acknowledges that user queries often embed multiple logical steps. Techniques such as query decomposition and sequential sub-question answering, popular in query rewriting and routing, have been demonstrated to enhance LLM performance by progressively building on intermediate results \cite{zhou2023leasttomostpromptingenablescomplex, yao2023reactsynergizingreasoningacting}.

\subsection{Boolean Retrieval}
In traditional search systems, methods for composing sophisticated search queries utilizing Boolean operators such as AND, OR, and NOT were proposed to achieve high recall and precision simultaneously. Although Boolean Retrieval demonstrated accuracy and high recall in search engines for specialized fields, its usability was ambiguous. This was because commands like AND, OR, and NOT were difficult to use without mastery. Furthermore, in search engines, there were often cases where natural language queries showed better performance than search queries using Boolean operators \cite{ref1}. However, retrieval recall and accuracy remain important for search platforms in knowledge and specialized fields, and these platforms often support Boolean operators and are centered on keyword-based searches.

\section{Methodology: Search like human with RAG (SHRAG)}

We will first provide an overview of the framework and then examine each component in detail below. For websites with an academic orientation in non English speaking regions, such as Wikipedia or ScienceON, there is a high probability that multiple languages may be intermixed within a single document. In particular, specific proper nouns, geographical names, and academic terms are frequently written concurrently with their original language notation (e.g., ``찰스 다윈 (Charles Darwin)''). Among these, English has the highest frequency of appearance; the background for this reason is that English is already established as the lingua franca of academia \cite{mauranen2003corpus}. Furthermore, in the ScienceON AI Challenge, questions were actually proposed in English that required finding answers in Korean materials, or conversely, English language answers had to be found for Korean language questions.
\begin{figure}[htbp]
  \centering
  \includegraphics[height=0.5\textheight]{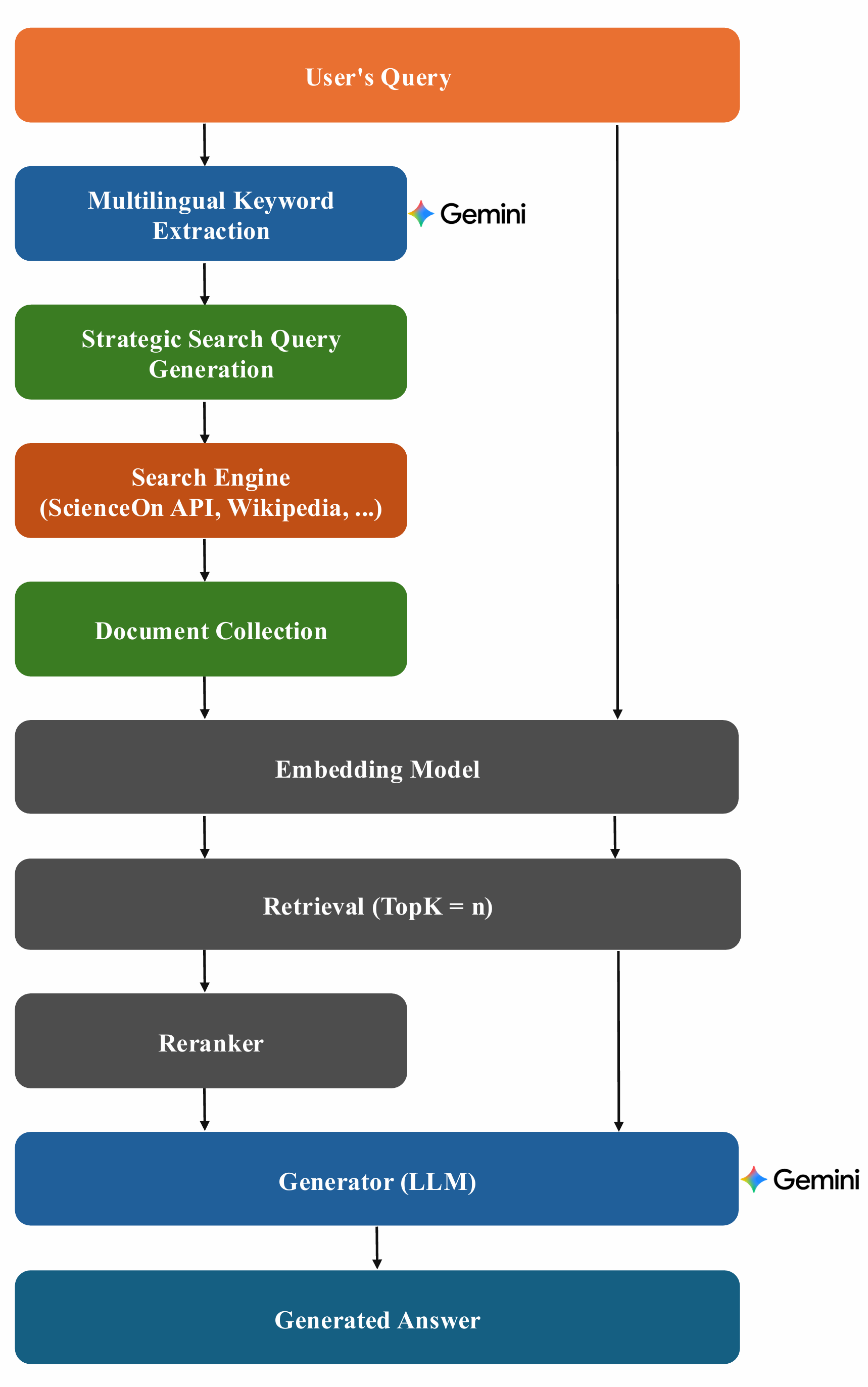}
  \caption{An illustration explaining the system's overall structure and components.}
\end{figure}
In this context, SHRAG responds to user queries through a process composed of the following stages: Multilingual Keyword Extraction, Strategic search query generation, document retrieval and collection, Multilingual Embedding, and retrieval and answer generation. Please refer to Appendix A for the detailed prompt implementation. We formalize the Search-Like-Human RAG (SHRAG) pipeline as a five-stage
mathematical framework that mimics the progressive narrowing behavior of a human searcher.
The overall flow is summarized in Algorithm~\ref{alg:shrag} (pseudocode) and
detailed mathematically below.

\subsubsection{Input and Output}
\begin{equation}
\begin{aligned}
\text{Input:} \quad & Q \in \mathcal{L}_{\text{query}} \quad
   (\text{Korean or English}) \\[4pt]
\text{Output:} \quad & A = (T, I, B) \quad
   (\text{Title}, \text{Introduction}, \text{Main Body})
\end{aligned}
\label{eq:input-output}
\end{equation}

\subsubsection{Step 1: Multilingual Keyword Extraction}
Let $\text{LLM}_{\text{ext}}(\cdot)$ be a large language model prompted to
return the top-$k$ keywords for a given language.
\begin{equation}
\begin{aligned}
K_{\text{en}} &= \text{LLM}_{\text{ext}}(Q,\ \text{lang}=\text{en},\ k), \\[4pt]
K_{\text{ko}} &= \text{LLM}_{\text{ext}}(Q,\ \text{lang}=\text{ko},\ k), \\[4pt]
K_{\text{word}} &= \text{SplitCompound}(K_{\text{en}} \cup K_{\text{ko}}), \\[4pt]
K &= \text{Rank}_{\text{imp}}(K_{\text{word}})[:10]
   = \{k_1, k_2, \dots, k_{10}\},
\end{aligned}
\label{eq:keyword-extraction}
\end{equation}
where $\text{Rank}_{\text{imp}}(\cdot)$ scores words by TF-IDF, LLM-based
importance, or entropy (implementation-specific).

\subsubsection{Step 2: Strategic Query Generation}
For each $n$ from 10 down to 1 we create a disjunctive query that becomes
progressively more specific:
\begin{equation*}
SQ_n = \bigvee_{i=1}^{n} k_i, \quad \forall n\in\{1,2,\dots,10\}.
\end{equation*}

\subsubsection{Step 3: Document Retrieval and Deduplication}
A standard search engine returns the top-10 hits for each query:
\begin{equation}
\begin{aligned}
D_n &= \text{SearchEngine}(SQ_n,\ \text{top}=10), \quad \forall n, \\[4pt]
D_n' &= \text{Filter}(D_n)
   \quad (\text{remove duplicates and incomplete entries}), \\[4pt]
D &= \bigcup_{n=1}^{10} D_n'.
\end{aligned}
\label{eq:retrieval}
\end{equation}

\subsubsection{Step 4: Multilingual Re-ranking}
We embed the query and every retrieved document with a multilingual GTE model:
\begin{equation}
\begin{aligned}
\mathbf{e}_Q &= \text{Embedding}(Q), \\[4pt]
\mathbf{e}_{d_i} &= \text{Embedding}(d_i),\quad \forall d_i\in D, \\[4pt]
s_i &= \cos(\mathbf{e}_Q,\mathbf{e}_{d_i}), \\[4pt]
D_{\text{top5}} &= \text{TopK}(D,\ s_i,\ 5).
\end{aligned}
\label{eq:reranking}
\end{equation}

\subsubsection{Step 5: Structured Answer Generation}
The final answer is produced by a generative LLM conditioned on the query
and the five most relevant documents:
\begin{equation}
A = \text{LLM}_{\text{gen}}\!\bigl(Q,\ D_{\text{top5}},\ \text{prompt}_{\text{structured}}\bigr),
\label{eq:generation}
\end{equation}
where $\text{prompt}_{\text{structured}}$ instructs the model to output
\emph{Title}, \emph{Introduction}, and \emph{Main Body}.
\subsection{Multilingual Keyword Extraction}
As previously mentioned, websites with an academic orientation in non-English-speaking regions have a high probability of containing English; additionally, it has been reported that performance and consistency improve when the prompt language matches the output language [Retrieval-augmented generation in multilingual settings]. Therefore, in SHRAG, we composed and executed prompts tailored to each respective language to extract keywords in English and keywords in a specified language (e.g., Korean) from an LLM based on the query. In this process, a range for the number of keywords to be generated is specified, and the keywords are retrieved in order of importance. The keywords generated from each language are then combined to create a Search Query Set. However, user queries can sometimes include multiple hops \cite{zhou2023leasttomostpromptingenablescomplex, yao2023reactsynergizingreasoningacting}. In this study, because the questions provided in the ScienceON AI Challenge were of the single-hop type, the design applying multi-hop is attached in Appendix B.
It was confirmed that compound keywords composed of two or more words (e.g., ``free textbook'' among the extracted keywords degrade search performance. Accordingly, all keywords were split by spaces and refined into single word keywords. A maximum of 10 keywords were used; including the split keywords, the top 10 keywords in order of importance were ultimately applied.

\begin{figure}
  \centering
  \includegraphics[width=\linewidth-80pt]{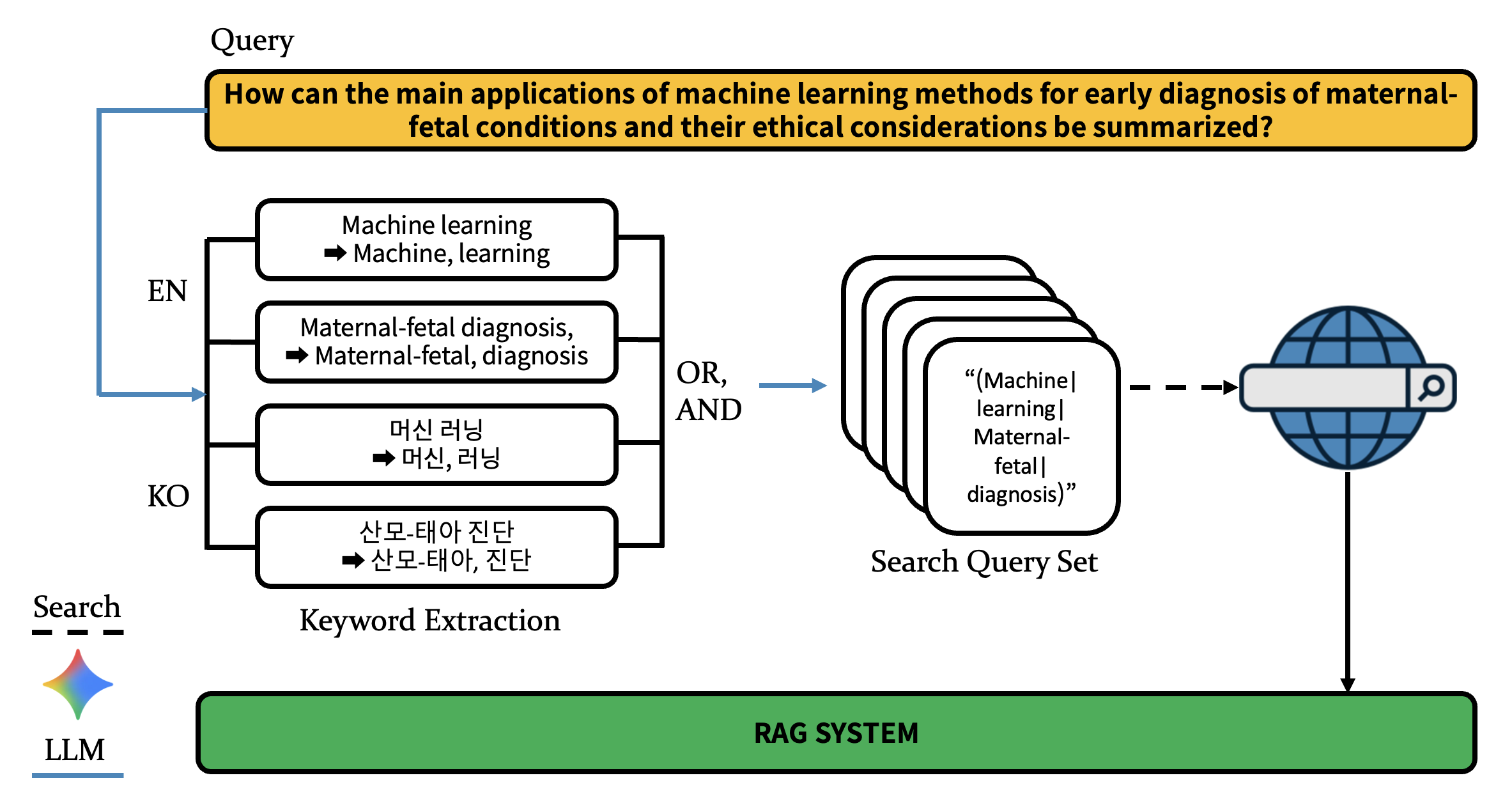}
  \caption{An illustration explaining Multilingual Keyword Extraction and Search Query Set generated by Strategic Search Query Generation.}
\end{figure}

\subsection{Strategic Search Query Generation}
Search queries were generated by combining the keywords extracted for each language with the OR operator. This aims to secure a broad range of potentially relevant documents, rather than acquiring only a few highly accurate ones. This approach is closely related to the characteristics of the Retrieval Augmented Generation (RAG) structure.
If search queries are precisely designed to target only specific documents, it becomes difficult to obtain meaningful answers even when using an RAG system should some key documents be omitted during the retrieval stage. Conversely, even if the search cost increases somewhat, securing diverse documents and including the relevant document within them allows RAG to retrieve the information and generate a more accurate and enriched response.
A comparative evaluation of OR based versus AND based query strategies is presented in the Experiments section. Queries incorporating the NOT operator were considered but excluded from this study, as their effective generation requires deep contextual reasoning by the LLM beyond the scope of current keyword based extraction. This direction is deferred to future work.
In the search query generation process, we started with the full keyword list and generated progressively reduced keyword combinations by removing keywords from the end of the list one by one. Because the keyword list is sorted by importance, this progressive reduction method ensures diversity in the search queries while maintaining the thematic consistency of the original query.

\subsection{Document Search and Collection}
A maximum of 10 documents was collected per Search Query via the search engine. For the collected documents, a final document set was constructed by performing duplicate document removal and verification of whether key information, such as paper abstracts, was missing. This procedure was applied iteratively to all Search Queries derived from a single Query.

\subsection{Multilingual Embedding}
The ScienceON AI Challenge includes a cross lingual query environment where Korean and English are mixed. To address this, this study adopted a multilingual embedding model. This model aims to improve cross language retrieval performance by effectively mapping the semantic representations between different languages into a common embedding space.
Model selection was performed by referencing the Massive Multilingual Text Embedding Benchmark (MMTEB) \cite{ref5}. The selection criteria were twofold:
First, the model must exhibit superior multilingual retrieval performance for the two languages included in the ScienceON AI Challenge queries, which are Korean and English. Second, it must support long text embedding functionality, capable of embedding long texts of 8,000 tokens or more at once.
Among the models that satisfied these criteria, the Snowflake and mGTE models, which showed excellent Korean–English multilingual retrieval performance, were selected as candidates \cite{ref6, ref12}. The Jina model also met the conditions but was excluded due to implementation constraints in the experimental environment \cite{ref13}.
Comparative experimental results on the candidate models indicated that the mGTE model computed the highest similarity between queries and the relevant documents. Consequently, this study ultimately adopted mGTE as the multilingual embedding model.

\begin{algorithm}[H]
\SetAlgoLined
\KwIn{User query $Q$ (in Korean or English)}
\KwOut{Final answer $A$}
\caption{Search like human with RAG (SHRAG)}
\label{alg:shrag}

\textbf{Step 1: Multilingual Keyword Extraction}\\
Extract top-$k$ English keywords $K_{\text{en}}$ from $Q$ using LLM\\
Extract top-$k$ target language (e.g., Korean) keywords $K_{\text{ko}}$ from $Q$ using LLM\\
Split compound keywords into single word units\\
Combine and rank keywords by importance $\rightarrow K = \{k_1, k_2, \dots, k_{10}\}$

\vspace{0.5em}
\textbf{Step 2: Strategic Search Query Generation}\\
\For{$n = 10$ \textbf{down to} $1$}{
    Generate search query $SQ_n = k_1 \vee k_2 \vee \dots \vee k_n$
}

\vspace{0.5em}
\textbf{Step 3: Document Search and Collection}\\
\For{each search query $SQ_n$}{
    Retrieve top 10 documents $D_n$ using search engine \
}
Remove duplicates and incomplete entries \\
Construct final document set $D = \bigcup_{n=1}^{10} D_n$

\vspace{0.5em}
\textbf{Step 4: Multilingual Embedding}\\
Encode $Q$ and all $D_i \in D$ using multilingual embedding model (mGTE)\\
Compute cosine similarity to select top 5 relevant documents $D_{\text{top5}}$

\vspace{0.5em}
\textbf{Step 5: Retrieval and Answer Generation}\\
Provide $\{Q, D_{\text{top5}}\}$ to LLM (Gemini 2.5 Flash)\\
Generate structured answer $A$ (Title, Introduction, Main Body)

\end{algorithm}

\subsection{Retrieval and answer generation}
Subsequently, the top five documents were retrieved using the retriever and provided to the LLM (Gemini 2.5 Flash) along with the query to generate the final answer. The prompt was designed to guide the model to produce responses that conform to the competition's required format, consisting of a title, introduction and main body.

\section{Experiments}

We measured the extent to which the documents collected during the keyword based document search stage included the relevant documents. This evaluation was based on the premise that, by leveraging the characteristics of the RAG structure, there is a possibility of generating accurate and reliable answers through the Retrieve process if the relevant documents are included among the searched documents. Furthermore, to evaluate the performance of the Boolean operators used during search query generation, we conducted a comparative experiment distinguishing between search queries generated using only the OR operator and those including the AND operator.
Additionally, to verify whether the proposed method exhibits consistent performance in a general environment without being dependent on a specific platform, we performed an identical experiment in a Wikipedia based search environment utilizing the MIRACL dataset.
The Large Language Models (LLMs) used in the experiment were Gemini 2.5 Flash and Gemini 2.5 Pro; a comparison of the two models revealed that Gemini 2.5 Flash showed the most superior performance relative to its processing time. Therefore, Gemini 2.5 Flash was used as the base model in the subsequent main experiments.

\section{Results}
\subsection{Evaluation with the ScienceON AI Challenge}
\subsubsection{Search Performance Evaluation Based on Operators}
When generating search queries using the AND operator, the queries were constructed by starting from the lower priority keywords in the importance ranking and progressively adding the specified number of higher priority keywords.
(e.g., if AND=1, search query=``free\textbar textbook\textbar mathematics$+$school'', if AND=2, search query=``free\textbar textbook$+$mathematics$+$school'')
The generated search queries were utilized in the experiment by varying the number of AND operators from 0, when only OR was used, to a maximum of 9, when only AND was used. Under the same condition, meaning the number of AND operators, document collection was performed for each of the 50 queries from the ScienceON AI Challenge, and each condition was executed 10 times to ensure experimental consistency. We have measured the proportion of answer documents that are included in search documents.

The experimental results, as shown in Figure 3, indicated that generating search queries using only the OR operator showed superior document collection performance compared to cases that included the AND operator.
Furthermore, search efficiency could be compared through the number of collected documents. As can be seen in Figure 3, search queries using only OR (0 ANDs) collected fewer than approximately 1,500 documents, whereas search queries containing AND collected approximately 2,000 or more documents. These results indicate that when the AND operator is included in the search queries, although a wider variety of documents may be collected, there is a tendency for document relevance to decrease. In other words, the AND operator is interpreted as overemphasizing the specific connected keywords, thereby leading to search results that deviate from the core topic.

\begin{figure}
  \centering
  \includegraphics[width=\linewidth-80pt]{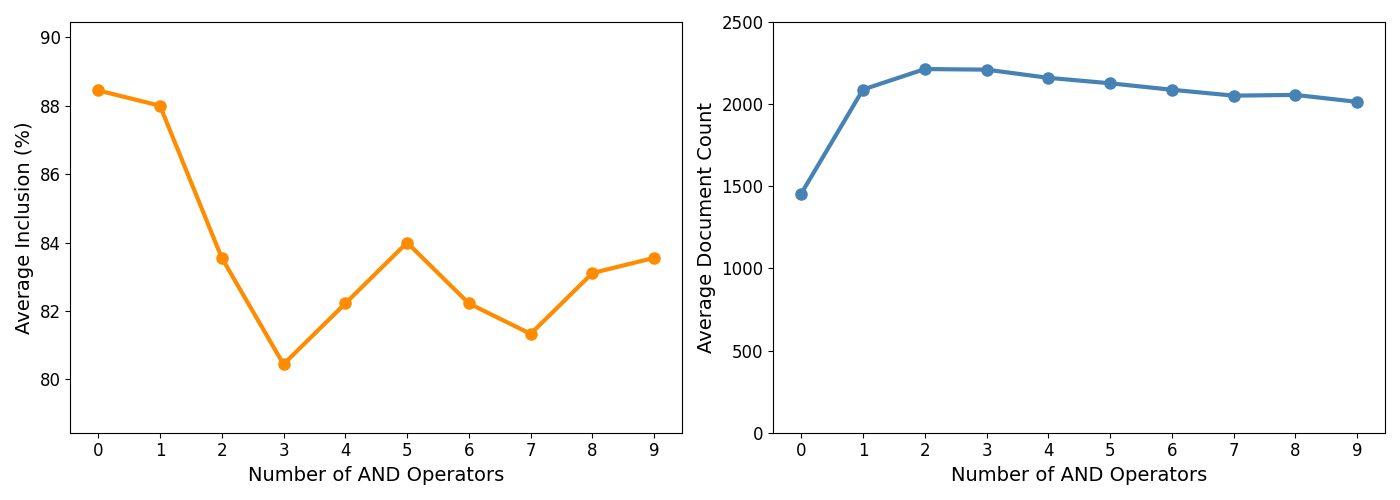}
  \caption{Left: The proportion of relevant documents that were successfully searched, categorized by the count of AND operators. Right: Average number of searched documents (over all queries) categorized by the count of AND operators.}
\end{figure}

\subsubsection{ScienceON AI Challenge Competition Query Analysis}

Furthermore, we submitted and evaluated results for both multi-hop and single-hop query assumptions during the competition; the resulting score difference was not significant. Human judges subsequently confirmed that the queries were almost exclusively single-hop, meaning answers could typically be found within a single paper. Therefore, assuming a single-hop context yielded marginally higher scores than applying a multi-hop framework.

\subsubsection{Winning at ScienceON AI Challenge competition}
Based on the results of the aforementioned analysis, we selected a multilingual search query expansion strategy using only the OR operator. Additionally, having confirmed that the majority of competition queries were single-hop, we did not split the stages of the original user query. This design choice optimized the system for retrieving answers from a single source. As a result, our framework achieved the highest overall performance on the criteria comprehensively evaluating Relevance of Answer to Ground Truth, Evidence Coverage, Retrieval Ratio, and Total Elapsed Time, ultimately winning the competition.

\subsection{Validation with the MIRACL Dataset}
Additional experiments were conducted to verify the generalization performance and effectiveness of the proposed framework. This experiment focused on `Evidence Coverage', one of the key evaluation metrics from the ScienceON Challenge. This metric evaluates whether the collected document set includes the ground truth document for a given query. Verifying this required a dataset that includes both queries and relevant documents, and provides the relevant documents in a searchable format on a public platform. As a dataset that fulfills these conditions, we utilized the MIRACL dataset, which contains various queries from Wikipedia and their corresponding relevant documents. 

For the experiment, a query set of 50 queries was constructed by randomly sampling 25 Korean queries and 25 English queries from the MIRACL dataset. The framework was applied to each query to retrieve approximately 15 documents from Wikipedia, and we measured whether the ground truth documents were included among them. 

We evaluate the search performance using the Query Success Rate (QSR), a metric defined to evaluate our framework on the MIRACL dataset. Since the MIRACL dataset provides one or more ground truth documents for each query, this metric measures the proportion of queries for which at least one relevant document is successfully retrieved. We judged this metric to be sufficient for our evaluation, as the retrieval of even a single ground truth document provides the necessary evidence for a RAG system to potentially generate the correct answer.

Let $Q$ be the set of all evaluation queries, with $|Q|$ representing the total number of queries. For each query $q_i \in Q$, we define two sets:
\begin{itemize}
    \item $A_i = \{a_{1i}, a_{2i}, \ldots, a_{ki}\}$: the set of $k_i$ known relevant document titles for the query $q_i$.
    \item $S_i$: the set of document titles retrieved by the search for the query $q_i$.
\end{itemize}

A query $q_i$ is considered successful if its retrieved set $S_i$ contains at least one relevant document (i.e., $A_i \cap S_i \neq \emptyset$).

The QSR is formally defined as the ratio of successful queries to the total number of queries:
\begin{equation}
\text{QSR} = \frac{|\{q_i \in Q \mid A_i \cap S_i \neq \emptyset \}|}{|Q|}*100
\label{eq:qsr} % (선택 사항) 본문에서 \ref{eq:qsr}로 참조할 때 사용
\end{equation}

As a result, it was found that the retrieved document set contained at least one ground truth document for 94\% of the total queries (94 QSR) . This demonstrates that SHRAG can effectively secure the necessary evidence documents required for generating comprehensive and accurate answers.
 
\begin{table*}
  \caption{Query Success Rate with MIRACL Dataset}
  \label{tab:relevant_ratio}
  \begin{tabular}{lcl}
    \toprule
    Query Language & QSR & Comments \\
    \midrule
    English & 100 & All queries contained at least one relevant document. \\
    Korean & 88 & Some queries did not yield relevant documents. \\
    English + Korean & 94 &The average QSR score. \\
    \bottomrule
  \end{tabular}
  
  \vspace{2mm}
  \small
\end{table*}

\section{Future Work}
The current framework operates by employing an LLM to extract keywords from a user's query, which are then reconfigured into search queries via a manually engineered combination logic. This approach has two primary limitations. First, it results in a rigid architecture that is tightly coupled to a specific search platform. Second, it is vulnerable to the lexical gap problem, where retrieval fails if the exact keywords from the query are not present in the document text. Our subsequent research will focus on addressing these constraints.

\section{Conclusion}
In this paper, we have proposed and validated SHRAG, a novel framework that leverages a Large Language Model to mimic sophisticated human search strategies. SHRAG's capabilities were empirically proven through its first place performance in the Kaggle-ScienceON AI Challenge, where it excelled in retrieval performance, evidence collection, and final answer quality. Its efficacy was further substantiated through additional verification with the MIRACL dataset, which reaffirmed its capacity to effectively retrieve a corpus containing the necessary information to answer a user's query.
A key strength of SHRAG lies in its plug and play architecture, which allows for high scalability in diverse downstream applications like multilingual tasks and can be deployed without any task specific fine tuning. Moreover, it demonstrates robust performance across both multilingual and monolingual scenarios, contingent on the availability of a powerful underlying search platform. Consequently, this research presents a new paradigm for RAG development. It suggests that the path to advancing RAG systems lies not only in the creation of superior embedding models or retrieval algorithms, but also in the strategic and intelligent utilization of existing search infrastructures.

\section*{Declaration on Generative AI}

During the preparation of this work, the authors used Gemini, Grok and ChatGPT in order to: translation, Grammar and spelling check. After using these tools/services, the authors reviewed and edited the content as needed and take full responsibility for the publication’s content. 

\begin{acknowledgments}
  Thank you for supporting the Institute of Information \& communications Technology Planning \& Evaluation (IITP) grant funded by the Korean government (MSIT) (RS-2024-00445087, No.2019-0-01842, Artificial Intelligence Graduate School Program (GIST)).
\end{acknowledgments}

\bibliography{sample-ceur}
% \printbibliography

%%%% If your work has an appendix, this is the place to put it.

\appendix
 
\section{Example Prompts}
\subsection{Example Prompt for Multilingual Keyword Extraction}
These are example prompts for Multilingual Keyword Extraction. Below show english and korean both with example results.
\input{prompta1}
\subsection{Example Prompt to Generate Answer for ScienceON AI Challenge Format }
This is example prompt for ScienceON AI Challenge Format. On ScienceON AI Challenge, Required Format of Korean and English was different.
\input{prompta2}
\subsection{Example Prompt for Query Decomposer}
\input{prompta3}

\section{Overall Structure with Query Decomposer for Multi-hop Queries}
\label{sec:overall-structure}

Figure~\ref{fig:overall-structure} illustrates the system's overall architecture, including the Query Decomposer module for handling multi-hop queries. The decomposer breaks down complex user queries into sub-questions, enabling step by step retrieval and reasoning.

% --- 이미지 + 캡션 강제 여기서 끝! ---
{%
\centering
\includegraphics[height=0.45\textheight, keepaspectratio]{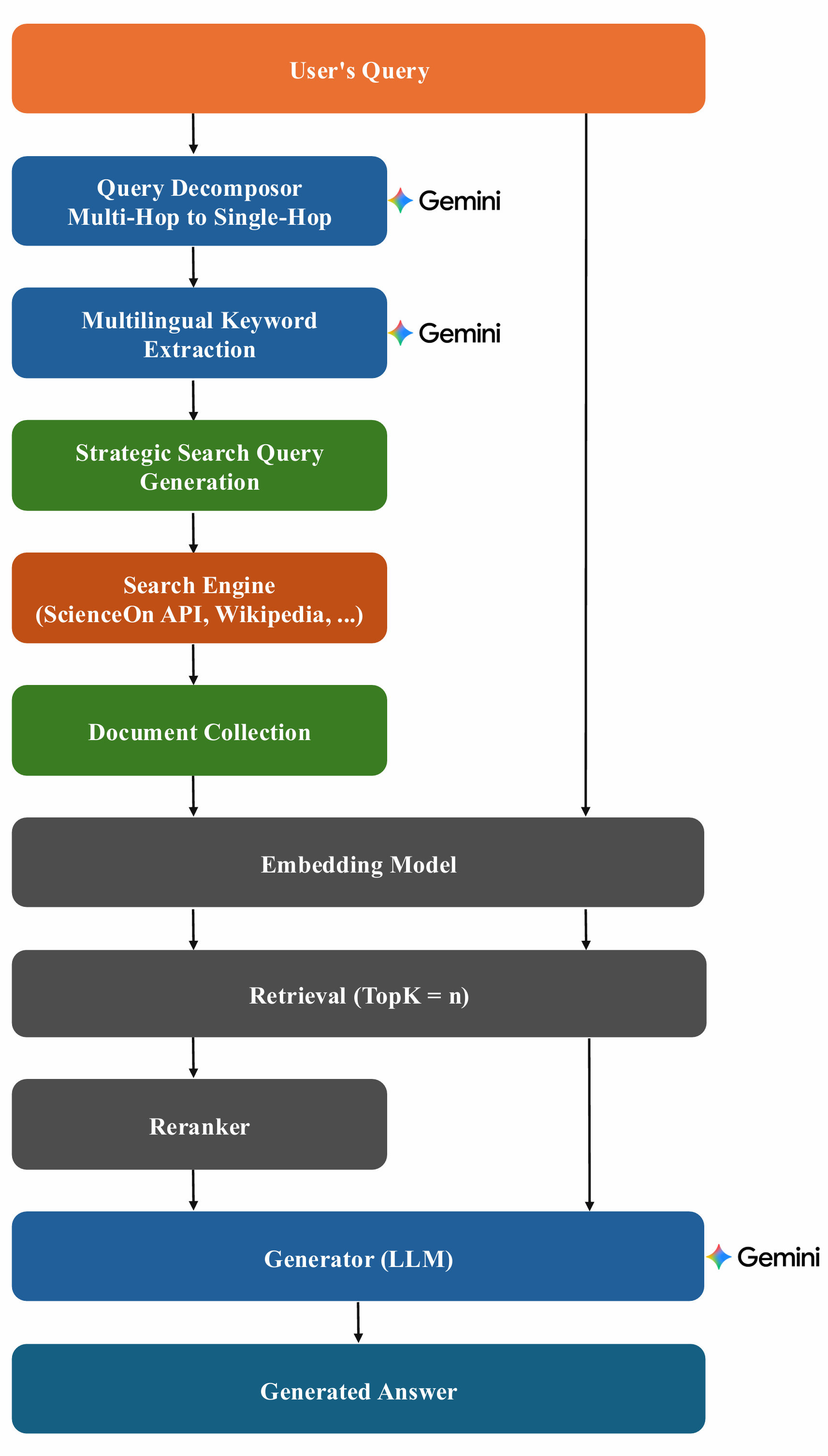}%
\vspace{8pt}%
\captionof{figure}{Overall system architecture with Query Decomposer for multi-hop query processing}%
\label{fig:overall-structure}%
\par
}%
% --- 끝 ---

\section{Online Resources}
\begin{itemize}
\item \href{https://github.com/omnyx2/SHRAG-A-Framework-for-Combining-Human-Inspired-Search-with-RAG}{GitHub: https://github.com/omnyx2/SHRAG-A-Framework-for-Combining-Human-Inspired-Search-with-RAG}
\end{itemize}

% \section{ Overall structure include Query Decomposer for multi-hop query }
% The Figure 4 is showing Query Overall structure with Query Decomposer. It decompose query with sub-questions.

% \begin{figure}[H]
% \centering
% \includegraphics[height=0.5\textheight, keepaspectratio]{overall1.png}%
% \caption{Overall system architecture with Query Decomposer for multi-hop query processing. The query is decomposed into sub-questions for sequential retrieval.}
% \label{fig:overall-structure}
% \end{figure}%

% \begin{itemize}
% \item \href{https://github.com/omnyx2/SHRAG-A-Framework-for-Combining-Human-Inspired-Search-with-RAG}{GitHub: https://github.com/omnyx2/SHRAG-A-Framework-for-Combining-Human-Inspired-Search-with-RAG}
% \end{itemize}

\end{document}

%% file: prompta1.tex
\begin{tcolorbox}[breakable, colback=blue!5!white, colframe=blue!75!black, title=Prompts for Multilingual Keyword Extraction]
\addvspace{10pt}
[English] \\
  
\addvspace{10pt}
  
You are an expert in extracting keywords for academic paper searches. From the given query, please extract the most important and useful keywords for the search in English.
 \\ \\
Query: ``\{query\}''\\
\\ \\
Requirements: \\
1. Extract in English keywords or number terms only \\
2. Prioritize technical terms and scientific jargon \\
3. In the case of abbreviations, use both the abbreviation and the \\full term as keywords \\
4. After extracting the keywords, list 2~5 in order of importance \\
5. List them in descending order of importance \\
\\ \\
Output Format: keyword1, keyword2, keyword3, keyword4 \\
\\ \\
Keywords: \\

\addvspace{10pt}

[Result Example] \\
Keywords: AI, Artificial Intelligence, Mathematics, Electronic textbook, Undergraduate course \\ 

\addvspace{10pt}
[Korean] \\

\addvspace{10pt}  

당신은 검색을 위한 키워드 추출 전문가입니다. 주어진 질문에서 가장 중요하고 검색에 유용한 키워드를 한국어로 추출해주세요. \\
\\
질문: ``\{query\}'' \\
\\
요구사항: \\ 
1. 한국어와 숫자 용어로 된 키워드만 추출하세요. \\
2. 전문용어와 기술용어를 우선적으로 선택하세요. \\
3. 축약어의 경우에는 축약어와 전체단어 모두 사용 키워드로 만드세요. \\
4. 2~5개 단어 추출하세요. \\ 
5. 중요도가 높은 순서대로 나열하세요. \\
\\
\\
출력 형식:키워드1, 키워드2, 키워드3, 키워드4 \\ 
\\
키워드:\\

\addvspace{10pt}

\addvspace{10pt}

[Result Example] \\
키워드: 인공지능 수학, AI, 전자 교과서, 학부 과정

\addvspace{10pt}

\end{tcolorbox}

%% file: prompta2.tex
\begin{tcolorbox}[breakable, colback=blue!5!white, colframe=blue!75!black, title=Generate Answer for ScienceOnChallenge Format]
You will be given a JSON object as a string which contains a series of related search queries and their retrieved documents (`hits'). Do not make answer from external knowledge. You must make answer inside of Context. \\
Your main task is to answer the specific `Question' provided below. Use the entire JSON data as context to formulate your answer, paying close attention to the `text' fields within the `hits' lists.\\
\\
The JSON data has a list of queries. The `original' query is the one you need to answer. The other queries are supplementary and provide additional context. \\ 
Do not use ``*'', ``-'' symbols. Please don't use ``*'' to emphasize words. Don't answer in markdown format just write it simple. \\ 
\\
If the `Question' is in Korean, format your answer in Korean as follows:\\ 
\#\#제목\#\#\\ 
\\
\#\#서론\#\#\\ 
\\
\#\#본론\#\# \\ 
\\
\#\#결론\#\# \\ 
\\
If the `Question' is in English, format your answer in English as follows, If English then Just write title inside of \#\#\{\{Title\}\}\#\#: \\ \\
\#\#\{\{Title\}\}\#\# \\
\\
\#\#Introduction\#\# \\ 
\\
\#\#Main Body\#\# \\ 
\\
\#\#Conclusion\#\# \\ 
\\
--- Context --- \\ 
\{context\} \\
\\
--- Question ---\\ 
\{query\} \\ 
\\

\addvspace{10pt}

[Result Example] \\
\#\#Summary of the Rationale and Structure of the Free Electronic Textbook for AI Mathematics\#\# \\

\#\#Introduction\#\# \\ 
This summary outlines the rationale and structure of a free electronic textbook, ``Basic Mathematics for Artificial Intelligence,'' designed for a one or two-semester undergraduate course. This textbook addresses the growing necessity of mathematical understanding for comprehending artificial intelligence (AI) and machine learning (ML) in the current era of advanced information and communication technologies. \\ 
\\ 
\#\#Main Body\#\# \\ 
The rationale behind developing this free electronic textbook stems from the recognition that understanding AI has become essential, and consequently, mathematical knowledge is now a mandatory prerequisite, not an option. Mathematicians are therefore tasked with explaining the mathematical foundations that enable AI technologies. To meet this need, a research team compiled the essential mathematical concepts required for understanding AI and machine learning into a curriculum suitable for a one or two-semester undergraduate course. \\ 

The structure of this educational initiative involves the authorship of the free electronic textbook, ``Basic Mathematics for Artificial Intelligence,'' which organizes these necessary mathematical concepts. Furthermore, courses based on this textbook have been established, targeting university students and graduate students from diverse academic backgrounds who are interested in the field of AI. The development process and operational examples of this textbook and its associated courses are shared to provide insights into its implementation. The textbook is accessible online at http://anonnymos.anonnymos.com \\
\\
\#\#Conclusion\#\# \\ 
In conclusion, the free electronic textbook ``Basic Mathematics for Artificial Intelligence'' was created out of the critical need for mathematical literacy in an AI-driven world. Its structure is tailored to provide a comprehensive, one or two-semester undergraduate curriculum covering the fundamental mathematical concepts essential for understanding AI and machine learning, making it accessible to a broad range of students.
 \\ 
\end{tcolorbox}

%% file: prompta3.tex
% ==============================
%  my_prompt_code.tex
%  (본문에서 \input{my_prompt_code} 로 불러오기)
% ==============================
\begin{tcolorbox}[
    breakable,
    enhanced jigsaw,
    colback=blue!5!white,
    colframe=blue!75!black,
    title=Generate Answer for ScienceOnChallenge Format,
]
\begin{lstlisting}[style=mypython]
DEFAULT_PROMPT = {
    "decompose": (
        """
You are an expert question decomposer.
Given a potentially multi-hop question (and optional context), produce the **minimal** set of independent, single-hop questions that, when answered, enable a solver to answer the original question.

Requirements:
- Each item must be answerable without external tools other than the given context and common knowledge.
- Avoid redundancy; keep the set as small as possible while sufficient.
- Prefer entity- and relation-focused questions.
- Do NOT answer anything. Only output the questions.
- Output strict JSON only, no extra text.

Return JSON with this schema:
{
  "single_hop_questions": ["q1", "q2", ...]
}
"""
    ).strip(),
    "chain": (
        """
You are an expert question planner.
Decompose the given multi-hop question into an **ordered** chain of single-hop questions such that each step leads naturally to the next and ultimately to the final answer.

Rules:
- Each question should be solvable on its own given prior steps and provided context.
- Keep the chain concise (3-6 steps typical). No answers, only questions.
- Output strict JSON only.

Schema:
{
  "steps": [
    {"index": 1, "question": "..."},
    {"index": 2, "question": "..."}
  ]
}
"""
    ).strip(),
    "rewrite": (
        """
You are an expert question rewriter.
Rewrite the given multi-hop question into a single, simpler **single-hop** question that preserves the original target and remains answerable (ideally with the provided context).

Rules:
- Keep core semantics and target unchanged.
- Remove multi-hop dependencies by substituting intermediate facts directly if they are present in the context. If not, reformulate to a single relation query.
- Output strict JSON only.

Schema:
{
  "single_hop_question": "..."
}
"""
    ).strip(),
}
\end{lstlisting}
\end{tcolorbox}